\def\year{2018}\relax
\documentclass[letterpaper]{article} 
\usepackage{aaai18}  
\usepackage{times}  
\usepackage{helvet}  
\usepackage{courier}  
\usepackage{url}  
\usepackage{graphicx}  

\usepackage{mathptmx}
\usepackage{amssymb}
\usepackage{amsmath}
\usepackage{subfigure}
\usepackage{hyphenat}
\usepackage{color}
\usepackage{textcomp}
\usepackage{enumitem}
\usepackage{verbatim}
\usepackage{algorithm}
\usepackage{algorithmic}
\usepackage{enumitem}
\usepackage{xspace}
\usepackage{subfigure}
\usepackage{adjustbox}

\usepackage{epigraph}

\usepackage[show]{chato-notes} 

\usepackage{cmm-greek}

\newcommand{\hide}[1]{}
\newcommand{\xhdr}[1]{\vspace{1.7mm}\noindent{{\bf #1.}}}

\newcommand{\papertitleOne}{How Constraints Affect Content:}
\newcommand{\papertitleTwo}{The Case of Twitter's Switch from 140 to 280 Characters}

\newcommand{\eg}{\textit{e.g.}\xspace}
\newcommand{\cf}{\textit{cf.}\xspace}

\newcommand{\vs}{\textit{vs.}\xspace}

\newcommand{\Secref}[1]{Sec.~\ref{#1}}

\newcommand{\Figref}[1]{Fig.~\ref{#1}}

\DeclareMathAlphabet{\mathcal}{OMS}{cmsy}{m}{n}

\begin{document}


\title{\papertitleOne\\\papertitleTwo%
}

\author{
Kristina Gligori\'c\\
EPFL\\
kristina.gligoric@epf\/l.ch
\And
Ashton Anderson\\
University of Toronto\\
ashton@cs.toronto.edu
\And
Robert West\\
EPFL\\
robert.west@epf\/l.ch
}

\maketitle

\begin{abstract}
It is often said that constraints affect creative production, both in terms of form and quality. Online social media platforms frequently impose constraints on the content that users can produce, limiting the range of possible contributions. Do these restrictions tend to push creators towards producing more or less successful content? How do creators adapt their contributions to fit the limits imposed by social media platforms?
To answer these questions, we conduct an observational study of a recent event: on November 7, 2017, Twitter changed the maximum allowable length of a tweet from 140 to 280 characters, thereby significantly altering its signature constraint.
In the first study of this switch, we compare tweets with nearly or exactly 140 characters before the change to tweets of the same length posted after the change. This setup enables us to characterize how users alter their tweets to fit the constraint and how this affects their tweets' success.
We find that in response to a length constraint, users write more tersely, use more abbreviations and contracted forms, and use fewer definite articles.
Also, although in general tweet success increases with length, we find initial evidence that tweets made to fit the 140\hyp character constraint tend to be more successful than similar\hyp length tweets written when the constraint was removed, suggesting that the length constraint improved tweet quality.

\end{abstract}

\section{Introduction}
\label{sec:intro}


\begin{flushright}
{\small
\textit{The enemy of art is the absence of limitations.}\,---Orson Welles
}
\end{flushright}

It is often thought that constraints affect both the form and quality of creative content. 
For example, there is anecdotal evidence across many domains that imposing length constraints can shape and improve the resulting writing: academic authors edit papers to fit a page limit, poets adhere to a prescribed verse form or rhyme scheme, and journalists edit articles to fit a word count limit \cite{McPhee2015}. More broadly, research in several fields---spanning product design \cite{creativity_thesis,moreau2005designing}, process management \cite{fritscher2009supporting}, and education \cite{hennessey1989effect}---suggests that having too much freedom can be paralyzing (epitomized by the feeling of staring at a blank sheet of paper), and that there is a sweet spot with just the right amount of constraints. 

Many social media platforms enforce length restrictions for posts, which is why the latter are frequently called \textit{microposts}.
For instance, at the time of writing,
Instagram captions were limited to 2,200 characters,
LinkedIn updates to 700 characters,
Pinterest pins to 500 characters,
and Twitter posts, \textit{tweets,} to 280 characters.
In the context of social media, a less restrictive character limit could conceivably be either positive or negative. 
In this work, we investigate the relationship between length constraints on post length and the form and quality of the resulting content. Do length constraints steer users towards editing their content in a useful way on social media, or does additional space allow for more creative and engaging content? 
Answering this question involves resolving two competing hypotheses:

\begin{enumerate}[label=\textbf{H\arabic*}, wide=0em, leftmargin=*]
    \item Imposing constraints has a positive effect on creativity, influencing users to create succinct content that is more likely to be appealing to others \cite{creativity_thesis}.
    \item Relaxing constraints provides users more space for expressing opinion and allows for more potentially interesting content. Also, longer posts occupy a larger portion of the update feeds displayed to users \cite{twitter_blog2}, further increasing engagement.
\end{enumerate}
%
%
Choosing the right
character limit is not trivial,
given these conflicting hypotheses.
Resolving them would not only help elucidate the relation between content and constraints; given the multi\hyp billion\hyp dollar economy depending on social media, the question is also of significant financial consequence.

The ideal way to answer this question would be through a randomized control study:
subject randomly sampled users to various character limits, and observe how the nature and success of their posts depends on the respective limit.
Unfortunately, however, only social\hyp media providers themselves have full control over their platforms, including the ability to conduct such A/B tests. 
Furthermore, even this ideal experiment might suffer from creating an ecosystem with differing content lengths with clashing norms and conventions.
Studying the effect of constraints is therefore difficult.

\xhdr{Twitter's November 2017 switch}
In order to circumvent these difficulties, we take advantage of a recent event to conduct an observational study of how length constraints affect microposts: on November 7, 2017, Twitter suddenly and unexpectedly increased its maximum tweet length from 140 to 280 characters.

The switch was reportedly introduced to allow users to express their thoughts without running out of characters, thus preventing them from finishing a thought \cite{twitter_blog1}. 
This change in length constraint---which we henceforth refer to as \textit{the switch}---constitutes an exogenous event that was most likely unexpected to most Twitter users, so we may reasonably assume that user behavior did not change in anticipation of this event. After carefully controlling for certain factors, differences between posts tweeted before \vs after the switch can inform us about the effect of different length constraints.

\xhdr{Research questions}
Given that our aim is to study the effect of length constraints on the style and success of content in online social media platforms, we seek to answer the following questions:
\begin{enumerate}[label=\textbf{RQ\arabic*}, wide=0em,  leftmargin=*]
    \item Do length constraints lead to characteristic changes in the writing style of posts?
    \item Do length constraints push users toward creating content that other users are more likely to engage with?
\end{enumerate}

\begin{figure}[t]
    \centering
        \includegraphics[width=0.45\columnwidth]{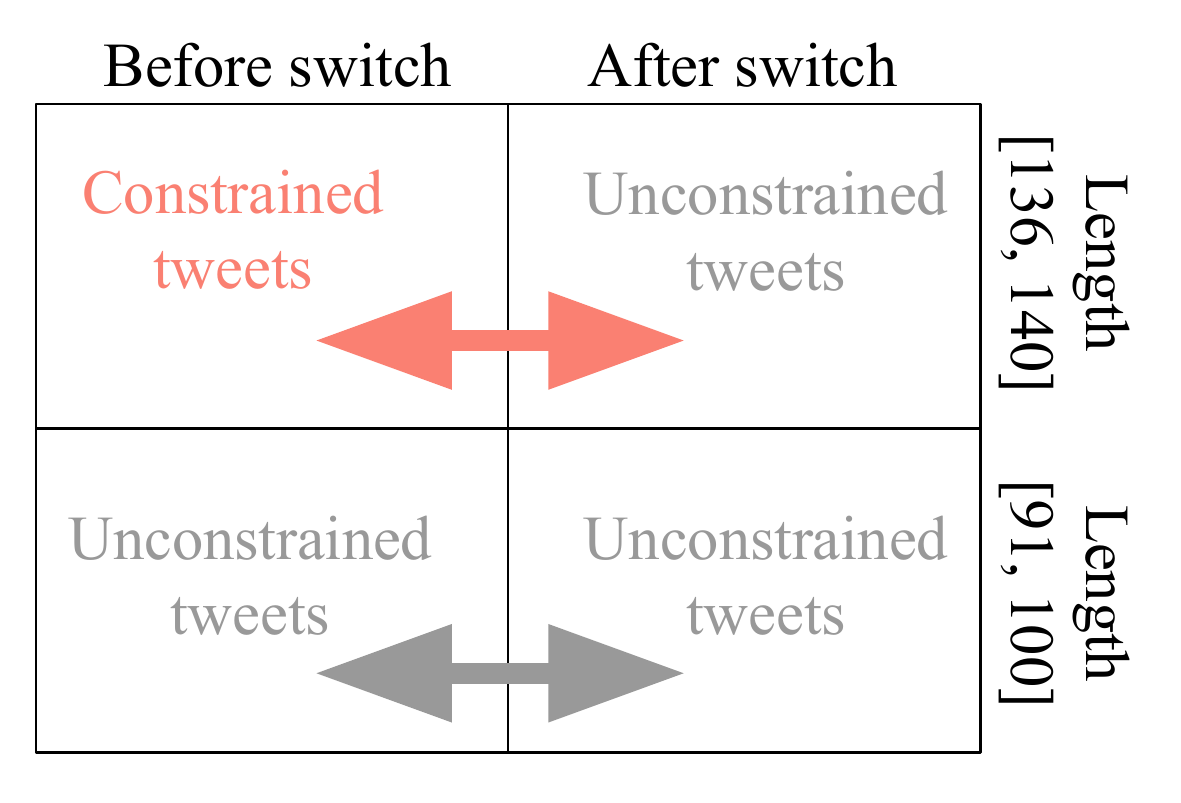}
        \includegraphics[width=0.45\columnwidth]{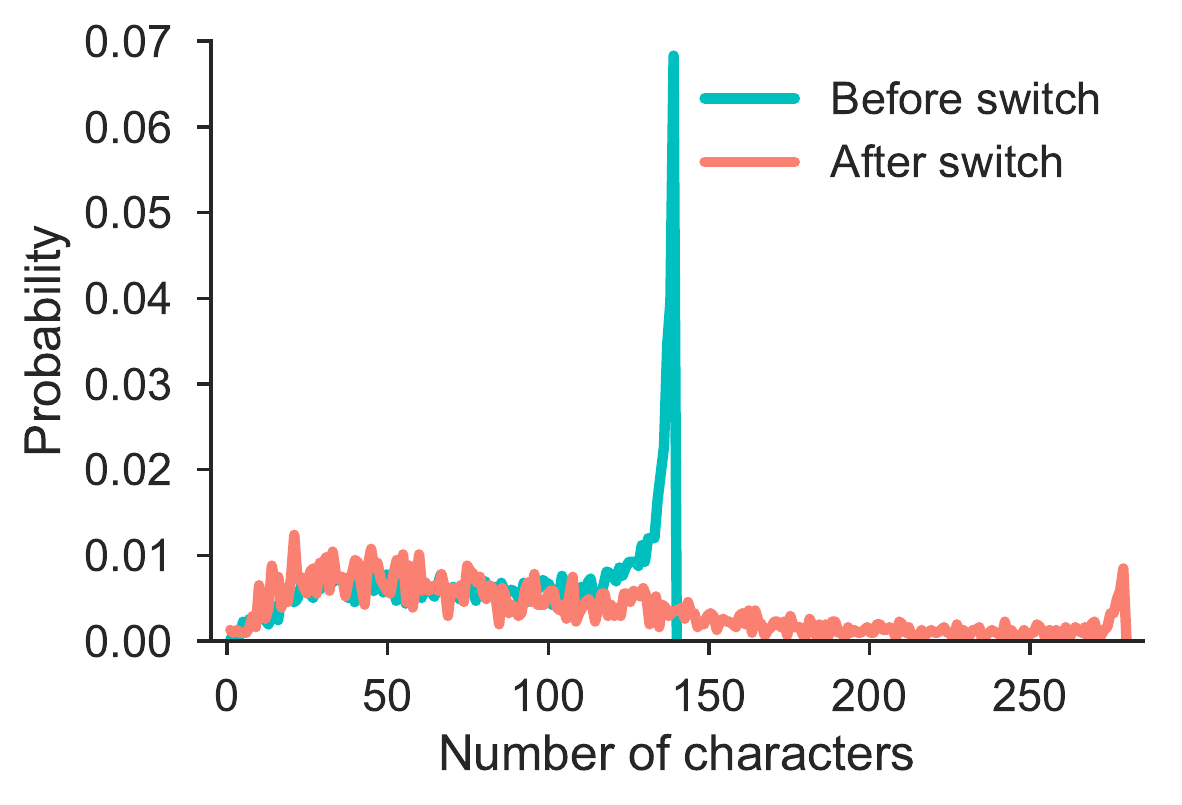}
    \vspace{-3mm}
    \caption{
    \textbf{Left:}~Schema of study setup.
    \textbf{Right:}~Histogram of tweet lengths, before and after the switch.}
    \label{fig:setup}
    \vspace{-3mm}
 \end{figure}

\section{Methodology%
\footnote{More details in the appendix.}
}
\label{sec:methods}

\begin{table*}[t]
\centering
\tiny
 \begin{tabular}{|l | c c | c c | c c |} 
 \hline
   &  \multicolumn{2}{|c|}{Matched for user} & \multicolumn{2}{|c|}{Matched for topic} &
   \multicolumn{2}{|c|}{Control study}
   \\
\hline
  Linguistic feature & Before & After & Before & After & Before & After\\
 \hline
Number of hashtags & 0.53 & 0.54 & --- & --- & 0.22 & 0.22\\
Number of emojis  & 0.25 & 0.25 & 0.37 & 0.29 & 0.31 & 0.32 \\
Number of abbreviations & \textbf{0.17} & \textbf{0.13} & 0.09 & 0.07 & 0.12 & 0.12 \\
Number of definite article  & \textbf{0.68} & \textbf{0.71} & 0.66 & 0.69 & 0.48 & 0.48\\
Number of indefinite articles  & 0.55  & 0.55 & 0.44 & 0.45 & 0.38 & 0.39\\ 
Number of \textit{and}  & \textbf{0.38} & \textbf{0.42} & \textbf{0.25} & \textbf{0.32} & 0.32 & 0.32\\
Number of \textit{\&}   & \textbf{0.16} & \textbf{0.11} & \textbf{0.18} & \textbf{0.14} & 0.03 & 0.03 \\
Number of missing spaces after punctuation  & 0.59 & 0.64 & 0.64 & 0.68 & 0.48 & 0.52\\
Number of auxiliary verbs and negations in long form  & 0.73 & 0.75 & 0.68 & 0.66 & 0.53 & 0.52\\
Number of auxiliary verbs and negations in contracted form  & \textbf{0.53} & \textbf{0.44} & \textbf{0.52} & \textbf{0.41} & \textbf{0.39} & \textbf{0.29}\\
Fraction of lexical words (nouns, verbs, adjectives, adverbs) & 56.45\% & 56.39\% & 55.44\% & 55.62\% & 57.34\% & 57.92\%\\
Lexical variation  & 0.9089 & 0.9081 & 0.9136 & 0.9121 & 0.9405 & 0.9410\\
Lexical sophistication  & 0.0014 & 0.0014 & 0.0014 & 0.0014 & 0.0013 &  0.0013\\
Number of words per sentence & 17.81 & 17.86 & 15.86 & 15.85 & 14.82 & 14.90 \\
 \hline
 \end{tabular}
  \hspace{0.4in}
 \begin{tabular}{|l | c c |} 
  \hline
 & \multicolumn{2}{|c|}{Matched for user} \\
 \hline
   & Before & After \\
 \hline
  \textit{I am}& 0.011 & 0.013\\ 
  \textit{I'm }& 0.057 & 0.051\\ 

  \textit{have}& 0.123 & 0.128\\ 
  \textit{'ve }& 0.033 & 0.026\\

  \textit{will}& 0.065 & 0.068\\ 
  \textit{'ll}& 0.027 & 0.021\\ 

  \textit{would}& 0.033 & 0.034 \\ 
  \textit{'d}& 0.014 & 0.011 \\

\hline
 
 \end{tabular}
\caption{
Results of linguistic analysis (RQ1).
\textbf{Left:}
Linguistic features averaged across tweets, before \vs after switch, when matching for users and topics (\Secref{sec:Matched studies}), and in control study (\Secref{sec:Control study}).
\textbf{Right:}
Mean number of occurrences per tweet of expanded and contracted word forms.
Bold values are significant with $p<0.00365$ according to Kolmogorov--Smirnov tests.}
\end{table*}

\begin{table*}[t]
\tiny
\centering
 \begin{tabular}{|l | c c | c c | c c | c c | c c | c c |} 
 \hline
 & \multicolumn{4}{|c|}{Matched for user} & \multicolumn{4}{|c|}{Matched for topic} &
 \multicolumn{4}{|c|}{Control study (tweet length $[91,100]$)} \\
\hline
  & \multicolumn{2}{|c|}{Retweets}
  & \multicolumn{2}{|c|}{Favorites}
  & \multicolumn{2}{|c|}{Retweets}
  & \multicolumn{2}{|c|}{Favorites}
  & \multicolumn{2}{|c|}{Retweets}
  & \multicolumn{2}{|c|}{Favorites}\\

  &  Before & After & Before & After & Before & After 
   &  Before & After & Before & After & Before & After\\
   \hline
  Probability of at least one & 21.61\% & 20.38\% & 45.50\%  & 45.52\% & 28.78\% & 29.94\% & 52.29\% & 52.17\% & 19.10\% & 18.72\% & 52.04\% & 52.94\% \\
  Quartiles (25\%, 50\%, 75\%) & (1, 2, 7) & (1, 2, 7) & (1, 3, 8) & (1, 3, 8) &  (1, 2, 5) & (1, 2, 4) & (1, 2, 6) & (1, 2, 6) & (1, 2, 5) & (1, 2, 5) & (1, 2, 6) & (1, 2, 6) \\
 \hline
 \end{tabular}
\caption{
Results of success analysis (RQ2).
\textbf{Top:}~Probability of receiving at least one engagement (retweet or favorite).
\textbf{Bottom:}~Quartiles of numbers of engagements (for tweets with at least one engagement).
}
\end{table*}

\begin{figure}[t]
    \centering
        \includegraphics[width=0.45\columnwidth]{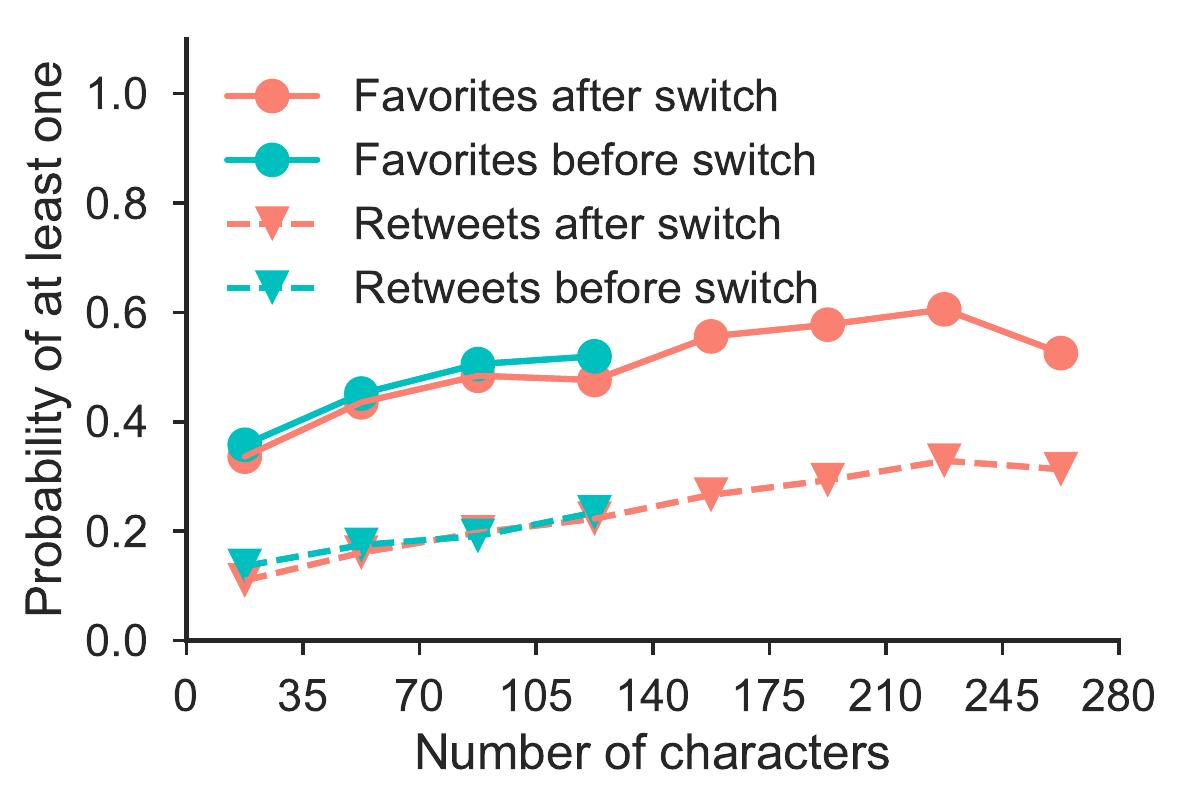}
        \includegraphics[width=0.45\columnwidth]{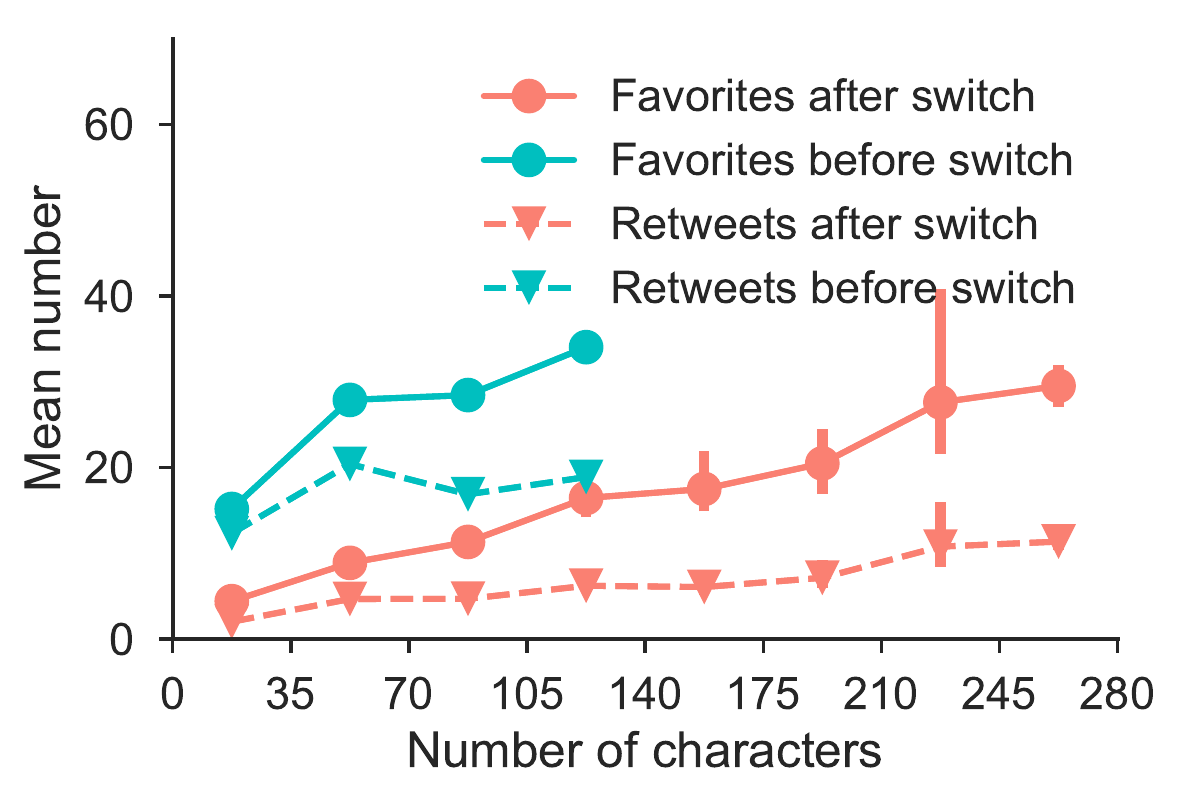}
    \vspace{-3mm}
    \caption{
    Engagement (retweets and favorites) as a function of length, with 95\% confidence intervals.
    \textbf{Left:} Probability of receiving at least one engagement.
    \textbf{Right:} Mean number of engagements (for tweets with at least one engagement).
    }
    \label{fig:results}
    \vspace{-3mm}
 \end{figure}

\xhdr{Research design}
In this paper, our main methodological contribution is a matched observational study design that allows us to use the exogenous shock to the Twitter ecosystem to investigate the relationship between length constraints and content form and quality, even in the absence of full experimental control of Twitter.
Before Twitter's switch from a 140-character limit to a 280-character limit, tweets that are nearly or exactly 140 characters in length are likely to have been explicitly ``squeezed'' by the user to comply with the character limit. 
After the switch, tweets of this length are less likely to have been affected by the character limit, since they are far short of the new  280-character maximum.
Our basic design is thus to compare various properties of tweets of a length between 136 and 140 characters%
\footnote{We use a 5-character range because it is not always possible to strictly optimize for the limit: omitting one word to meet the constraint may result in a tweet of fewer than exactly 140 characters.
}
just before \vs right after the switch (\Figref{fig:setup}, left).
The major difference being the presence \vs absence of the 140\hyp character limit, comparing tweets of this length from before \vs after the switch (corresponding to the upper, red arrow in \Figref{fig:setup}, left) lets us isolate the impact of the character limit. By comparing content written very close in time, we minimize the likelihood of external factors changing
in the time during which
our studied tweets were written.
In a control study, we also compare tweets with 91 to 100 characters before \vs after the switch, as these tweets are unaffected by the character limit during both time periods (lower, gray arrow in \Figref{fig:setup}, left).

\xhdr{Data}
In order to execute this research design, one would ideally analyze all tweets in our character range of interest that were authored right before and right after the switch. In lieu of access to complete Twitter data, we approximate this ideal.
We first collect tweets from the 1\% sample Twitter supplies via its \textit{Spritzer API,} for the time between April and June 2017. From this set of tweets we sample 100K users, where the probability of being sampled is proportional to the number of 140-character-long tweets they posted. This ensures that we are likely to sample users who disproportionately generate tweets for which the 140-character constraint is relevant. We proceed to collect these users' timelines using the Twitter \textit{user timeline API,} gathering up to the 3,200 most recent tweets per user. We simultaneously keep collecting new tweets these users post. All tweets we analyze were posted between November 2017 and January 2018.%
\footnote{Data available at \url{https://github.com/epfl-dlab/140_to_280}}

\xhdr{Controlling for users and topics}
With the goal of mitigating potential confounds in our comparison of tweets before \vs after the switch, we perform two analyses.
First, we control for users
by including, for each user, the same number of tweets written before the switch as after the switch (but this number may vary across users).
As we are interested in matching tweets actually written by a particular user, we do not consider retweets of other users' tweets.
Second, we also conduct an analysis in which we control for topics by including, for each unique set of hashtags, the same number of tweets before as after the switch with that set of hashtags.

\xhdr{Controlling for temporal effects}
By design, we compare tweets posted before and after the switch, thus introducing a time gap between the observed content.
This raises two problems:
first, retweet and favorite counts depend on how much time has passed since posting;
and second, one might observe differences in the number of retweets and favorites as a side effect of a user's becoming more popular with time.
We address the first issue by making sure to collect a tweet at least 48 hours after it was posted, since by that time 99.99\% of retweets happen \cite{retweets_saturation}.
The second issue is ruled out by making sure that the number of a user's followers did not increase by more than 1\% between the user's first and last tweets in our dataset, when controlling for users.

\xhdr{Constraint consistency over time}
We ensure that all content we analyze was generated under the same character counting policy. Additionally, we are primarily interested in how \textit{text} is altered under a particular restriction. We thus limit ourselves to tweets not containing URLs and @ replies. 

\xhdr{Pagination} Twitter users have long been working around the 140-character limit by splitting a long piece of text into a sequence of length-compliant tweets. These tweet threads are usually annotated with the position of the tweet in its sequence, \eg, \textit{2/3}. They are usually not intentionally altered in order to fit the character limit requirement using mechanisms we are interested in. Thus, we disregard such tweets.

\section{Results}
\label{sec:results}
\subsection{General impact of length constraints}

We begin with an analysis of the general impact of tweet length limits. We follow a data-driven approach and analyze 4M tweets written before, and 1.9M written after, the switch. Tweets are filtered according to our overall study setup (\Secref{sec:methods}).

\Figref{fig:setup} (right) contains a histogram of tweet lengths before and after the switch.
The spike at exactly 140 characters before the switch is indicative of users dealing with the character limit constraint. Additionally, the fact that there is no corresponding spike after the switch reflects the fact that the spike was indeed induced by the constraint---once it is lifted, 140 characters is an unexceptional tweet length.

\Figref{fig:results} presents the probability of obtaining at least one engagement (retweet or favorite) and the average number of engagements (given at least one engagement), as a function of tweet length.
We observe that, both before and after the switch, longer tweets are more successful on average, as captured by the number of retweets and favorites. This demonstrates the existence of a length effect consistent with hypothesis H2 (\Secref{sec:intro}).
This finding is, however, subject to several potential confounds; \eg, tweet length might correlate with the importance of what the user has to say, or with the popularity of the user herself, which might in turn be the real cause for the success of the respective tweet.
Our matched studies, described next, address these confounds.

\subsection{Matched studies}
\label{sec:Matched studies}

We compare tweets of length $[136,140]$ posted before, with tweets of a length in the same range posted after, the switch.

\xhdr{Linguistic analysis (RQ1)}
We are interested in linguistic aspects that might conceivably be altered in an attempt to meet length constraints: hashtags, emojis, abbreviations and acronyms used in the Twitter community \cite{twitter_jargon1}, articles, conjunctions, spaces after punctuation, and auxiliary verbs in their long, contracted, or incorrectly contracted forms.
We also evaluate stylistic features of text, including the frequency of lexical words,
lexical variation, and sophistication as defined by \citeauthor{Lu2010TheRO} (\citeyear{Lu2010TheRO}).
Moreover, we compute the number of words per sentence, an indicator of readability.
Here,
we are not interested in content that fails to elicit any engagement at all, as it is unlikely to reflect the optimizing mechanisms of interest.
We thus keep only tweets that have received at least one retweet or favorite, resulting in 12K (4K) tweet pairs when matching for users (topics).

Matching for users and for topics, respectively, we compare the above linguistic features before \vs after the switch using Kolmogorov--Smirnov significance tests. Given that we test multiple hypotheses (one per feature), the significance threshold is adjusted via Dunn--\v{S}id\'ak correction \cite{abdi2007bonferroni}, where the adjusted significance threshold is calculated as $1 - (1-\alpha)^{1/m}$.
We set $\alpha = 0.05$ and $m = 14$, resulting in an adjusted significance threshold of 0.00365.

Whether we control for users or for topics, constrained tweets systematically differ in linguistic features (Table 1). They contain fewer hashtags, fewer articles, and more words in their short forms: more abbreviations, more auxiliary verbs in their contracted forms, more \textit{\&}, and fewer \textit{and}.

These discrepancies are indicative of edits taking place given a 140-character limit constraint. We do not observe more missing spaces after punctuation in constrained tweets. This signals that users are unwilling to use punctuation in an incorrect way in order to fit a constraint.

\xhdr{Success analysis (RQ2)}
When quantifying engagement, we evaluate how much other users engage with a tweet in terms retweets and favorites. Controlling for users and for topics, respectively, we evaluate success before \vs after the switch. The results of this study are presented in Table 2. We see that, when controlling for users, constrained tweets are 6.0\% (relative) more likely to elicit engagement in the form of at least one retweet, compared to unconstrained tweets.
Note that the trend is reversed when controlling for topics (instead of users), but the difference is smaller ($-3.9$\% relative), and controlling for users is arguably a stricter criterion than controlling for topics, which, taken together, indicates that length restrictions slightly improve the success of tweets, in line with hypothesis H1 (\Secref{sec:intro}).

\subsection{Control study for ruling out confounds} 
\label{sec:Control study}

As our analysis compares two different time periods to each other, one might object that the observed stylistic changes were in fact caused by a community\hyp wide drift of norms between the two time periods.
Although unlikely given the condensed time frame, we aim to rule out this alternative explanation with a control study in which we repeat our analysis (matching for users; 9K tweet pairs), but this time comparing tweets of a length much below the 140\hyp character limit (91 to 100 characters) before \vs after the switch
(\cf\ \Secref{sec:methods}).

Such short tweets are unlikely to have been optimized for the length constraint. As a consequence, if any differences are observed for $[91,100]$ tweets,
the same mechanisms might be at work for $[136,140]$ tweets, and it might be these mechanisms---rather than the 140\hyp character constraint---that also cause the differences among $[136,140]$ tweets.
However, we observe only one statistically significant difference in linguistic features for the $[91,100]$ range (Table 1), hinting at the length constraint as the cause for the other features.

As for success, we observe that $[91,100]$ tweets posted before the change are 2.0\% (relative) more likely to be retweeted (Table 2), a much smaller difference than for $[136,140]$ tweets (6.0\% relative),
which we interpret as support for our above finding that length constraints  slightly increase tweet success.
However, more studies are necessary to confirm these findings, ideally with more data.

\section{Discussion}
\label{sec:discussion}
In this paper, we address the question of how users adapt their contributions to fit the limits imposed by social media platforms and whether these restrictions tend to push users towards producing more or less successful content. We develop a matched observational methodology that involves comparing carefully selected sets of tweets posted just before \vs just after the switch.

We find that tweets constrained by a 140\hyp character limit contain fewer hashtags and definite articles,
and more words in abridged form: more abbreviations, more contracted auxiliary verbs, more \textit{\&,} and fewer \textit{and}, implying that, when subject to a length constraint, users write more tersely. 

We also found initial evidence that length\hyp constrained tweets are slightly more successful in terms of the engagement they receive from other users.
However, future work needs to develop a better understanding of this phenomenon and determine whether the observed findings are robust.
We anticipate that collecting more data from the time after the relaxation of the constraint will be helpful to this end.

In our studies, we assume that tweets in the $[136,140]$ range posted after the switch are not constrained.
We cannot rule out the existence of users adhering to the old constraint despite the switch (\eg, out of sheer nostalgia, or because they are bots generating tweets via scripts written before the switch), but in the absence of such hypothetical users, the amplitude of the observed effects would be even larger.

\xhdr{Future work}
This paper presents the first in what we hope will be a series of studies regarding the impact of Twitter's switch from 140 to 280 characters.
In particular, our observational studies should be followed by experimental studies, in order to corroborate the initial findings presented here.
Also, our results are from right around the switch, when things were potentially still in flux.
It would be interesting to revisit our findings in the future once the Twitter ecosystem has settled into a new steady state.
Finally, we hope that future work will generalize our findings to constrained content production beyond Twitter.

{
\small
\raggedright
\bibliographystyle{aaai}
\bibliography{bibliography}
}

\appendix
\section{Appendix}
\label{sec:appendix}
The first four pages of this paper were published in the proceedings of ICWSM 2018.
In this appendix, which did not appear in the proceedings, we provide additional details about the implementation of our observational study and make some additional remarks.

\subsection{Implementation details}

\xhdr{Controlling for users and topics}
In the main studies, we build pairs of tweets posted by the same user or containing the same set of hashtags.
Note that a single user might post multiple such pairs of tweets. When matching, we choose the tweets such that the time distance from the switch is minimized: before the change, we keep the most recent tweet, and after the change, we keep the least recent tweet. As a result, selected pairs of tweets are tightly distributed around the switch. Three quarters of matched tweets posted before the switch are posted after October 9, and the median is October 26. In the case of tweets posted after the switch, three quarters are posted before December 9, and the median is November 23.

\xhdr{Constraint consistency over time}
Twitter has been gradually distancing itself from the original hard limit of 140 characters. First, they stopped counting media (including URLs and quoted tweets) and @ replies toward the character count, with these becoming metadata instead. However, after relaxing the character limit, such content is now counted again.
Therefore, in order to study the impact of the exact identical constraint during all time periods, we exclude tweets containing URLs, quoted tweets, and @ replies altogether when sampling tweets for our studies.
In order to rule out any further potential variations of the counting policy over time, and since it is impossible to determine the version of the application that was used to send a given tweet, we moreover consider only tweets that were posted using the Twitter web client. Furthermore, since the character limit is not equal for all languages \cite{twitter_blog2} and in order to maintain a meaningful interpretation of lexical features, we limit ourselves to tweets in English.

\xhdr{Pagination}
In our analysis, we detect and disregard paginated tweets in the following way. From the gathered data, we detect 1,722 instances of \textit{pagination} by detecting consecutive tweets posted by the same user containing a label in form of \textit{position of a tweet in a sequence/total number of tweets in a sequence}, \eg, \textit{1/3, 2/3, 3/3}. We infer that pagination is relatively rarely used, as less than 0.1\% of all tweets in our dataset are paginated. Furthermore, 90\% of subsequent paginated tweets are posted within four minutes of each other. In the subsequent analysis, we disregard tweets the same user posts within this time frame, but we observe that varying the time frame, and thus the percentage of detected paginations, does not qualitatively affect our findings.


\subsection{Note on a user interface update}
Alongside the change of the character limit, Twitter updated its user interface such that it no longer counts down the number of characters as the user types. Instead, a circle fills in as the limit is approached. As a consequence, the user is not aware of the exact number of characters left until only 20 (out of the total of 280) remain. As this change was made contemporaneously with the switch from 140 to 280 characters, we cannot rule out the possibility that our results are affected by this user interface change.
However, regardless of whether users see an exact character count or a half\hyp filled circle, it would likely be clear that tweets in the $[136,140]$ range that we study are not near the 280-character limit after the switch.
Moreover, by being shown a circle instead of an exact character count, users after the switch cannot realize how close exactly they are to the deprecated 140\hyp character limit and whether they have passed it, which effectively rules out optimization for the old 140\hyp character limit after the switch due to nostalgia or hipsterdom.
For the above reasons, we argue it is reasonable to assume that the change from displaying an exact count to displaying a circle has a negligible affect on our results.

\subsection{Final remark about engagement}
We also make an interesting observation with respect to the relation between tweet length and other users' engagement with a tweet:
\Figref{fig:results} shows that the success of tweets is strongly correlated with their length, both before and after the switch,
but also that tweets of a length up to 140 characters posted after the switch are vastly less popular than tweets of the same length posted before the switch.
Indeed, it seems that the total amount of engagement caused by the entirety of all tweets stays constant before \vs after the switch---it is merely distributed over a wider range of lengths after the switch.
It is interesting to see Twitter's switch in this light, given that it was officially motivated by the claim that tweets longer than 140 characters lead to more engagement \cite{twitter_blog2}.

\end{document}